\documentclass{revtex4}

\usepackage{graphicx}
\begin{document}
\bibliographystyle{revtex}


\title{A Single Shot, Sub-picosecond Beam Bunch
Characterization with Electro-optic Techniques}



\author{Y.K. Semertzidis, R. Burns, V. Castillo, R. Larsen,
D.M. Lazarus, D. Nikas, \\ 
C. Ozben, T. Srinivasan-Rao, A. Stillman, and T. Tsang}
\affiliation{Brookhaven National Laboratory, Upton, New York 11973, USA}

\author{L. Kowalski}
\affiliation{Montclair State University, Upper Montclair, NJ 07043, USA}


\date{\today}
\def\ni{\noindent}
\def\st{\scriptstyle}
\def\sst{\scriptscriptstyle}
\def\mco{\multicolumn}
\def\epp{\epsilon^{\prime}}
\def\vep{\varepsilon}
\def\ra{\rightarrow}
\def\ppg{\pi^+\pi^-\gamma}
\def\vp{{\bf p}}
\def\ko{K^0}
\def\um{\rm  \mu  m}
\def\us{\rm  \mu  s}
\def\kb{\bar{K^0}}
\def\al{\alpha}
\def\ab{\bar{\alpha}}
\def\be{\begin{equation}}
\def\ee{\end{equation}}
\def\bea{\begin{eqnarray}}
\def\eea{\end{eqnarray}}
\def\oma{\omega}
\def\CPbar{\hbox{{\rm CP}\hskip-1.80em{/}}}
\def\ecm{\rm \, e \cdot cm}

\begin{abstract}

In the past decade, the bunch lengths of electrons in accelerators have 
decreased dramatically to the range of a few
picoseconds~\cite{Uesaka94,Trotz97}.   
Measurement of the length as well as the longitudinal profile of these short 
bunches have been a topic of research in a number of
institutions~\cite{Uesaka97,Liu97,Hutchins00}. 
One of the techniques uses the electric field induced by the passage of 
electrons in the vicinity of a birefringent crystal to change its optical 
characteristics. Well-established electro-optic techniques can then be used 
to measure the temporal characteristics of the electron bunch.  In
this paper we  
present a novel, non-invasive, single-shot approach to improve the 
resolution to tens of femtoseconds so that sub-millimeter bunch length can be 
measured.

\end{abstract}

\maketitle

\section{Introduction}

The development of very short beam pulses is 
a crucial part of the effort to achieve the
desired luminosity for the future accelerators which are being
 designed to bring us to the next
energy frontier and for SASE for FEL.  The Lorentz cone of an
ultra-relativistic charged particle  
beam bunch (CPB) has a very small opening angle~\cite{jackson} and therefore
the amplitude of the generated
electric field has a longitudinal profile very close to the longitudinal
charge distribution of the beam.  One way to measure the produced electric
field in a non-invasive way is the electro-optic effect (Pockels
effect) using
 an electro-optic 
crystal as the electric field sensor.  
A polarized laser light goes
through the electro-optic  
crystal and the E-field from the CPB induces an ellipticity in the
polarization state of the laser light.  Ellipsometer techniques are used to
 analyze the laser light polarization state.  The first
electro-optic detection of  charged particle beams have been reported in 
references~\cite{pac99,yan00,fitch01,semertzidis00,tsang01}. 

\bigskip

The techniques that are used to read out the ellipsometers are:

\begin{enumerate}

\item{} The  pump-probe method where a short laser pulse is used 
with a determined time delay between the electron beam and the laser
pulse.  The 
time delay is varied so different parts of the electron beam can be probed.
This approach can in principle give the average longitudinal beam profile of 
many electron pulses~\cite{yan00,fitch01}.  

The time resolution of this
method is determined by the appropriate combination of the
laser pulse length, the laser pulse size in the crystal and the crystal
length ($CL$) along the laser propagation direction.  The latter is
equal to $\sigma_{cl} = CL / (c/n)$ with $c$ the 
speed of light and $n\approx 2.2$ 
the index of refraction for the $\rm Li Nb O_3$ crystal at the laser  
frequency.  For a crystal length of 1~mm,  $\sigma_{cl} \approx
7.5$~ps.
If the CPB length is shorter than the crystal length, the time
resolution would be equal to the CPB pulse length.  If the laser
pulse and the CPB move in the same direction then the pulse
broadening equals $\sigma_{cl} = CL (1-1/n) / c \approx 1.8$~ps for
the above example which is consistent with the resolution reported in
reference~\cite{yan00}.  
In order to reach sub-picosecond time resolution the crystal length needs
to be less than $100 \, \mu \rm m$.  Crystals with $100 \, \mu \rm m$
or smaller 
lengths are already commercially available.

\item{} The single shot technique where the light after the ellipsometer is
either (a) read by a streak camera~\cite{dimitri01}, or (b) the laser light is
stretched after it goes through the crystal so 
that the signal can become slow enough to be read out using conventional 
electronics~\cite{Hutchins00}
or (c) by measuring the laser pulse spectra with and without the presence
of an electron pulse~\cite{wilke01}.  The
advantage of the  
single shot method is that it can, in principle, provide information
on the longitudinal beam profile on a single pulse basis.  

The time resolution of this approach is limited by the laser pulse size in
the crystal and the crystal length along the laser propagation.  

\end{enumerate}

 We report a recent development~\cite{triveni01} based  on the
electro-optic effect which is capable of reaching sub-picosecond time
resolution down, in principle, to a few fs  in the single shot mode.

\section{Theory}

Let us consider an electron beam bunch of charge density $\sigma (x,
y)$, and bunch  
length $l$, focused to a sheet beam with transverse dimension 
$D$. Let this  
relativistic charged particle beam move along $x$ axis, the length of a 
birefringent crystal. The electric field experienced by the
crystal at a distance r from the  
electron beam, due to the charge $\sigma (x, y) dx dy$ can be written as

\be
{dE_z} = (\gamma / 4 \pi \epsilon_0) \sigma(x,y) \, dy \, dx \, /\epsilon r^2
\ee

\ni where $\epsilon $ is the dielectric constant of the crystal in Z
direction and  
$\gamma $ is the relativistic Lorentz factor. This field is present at this 
location for the time $dt$ taken by this charge to traverse the
distance $dx$. A  
polarized laser beam propagating along the y-axis inside this birefringent 
crystal would then experience this field over a distance $dL=dx/n$
where n is the refractive index of the crystal, along the direction of  
propagation at the laser frequency. The phase retardation experienced 
between the two orthogonal polarization components (z and x) of the laser 
beam $d \Gamma (t) $ is:

\be
d \Gamma(t) = \kappa (2 \pi / \lambda)  \, dL \, dE_z(t)
\ee

\ni where $\kappa $ is the electro-optic coefficient and $\lambda $ is 
the wavelength of the laser beam. The total retardation is obtained by 
integrating over the entire charge distribution, the time taken by the
laser to cross the  
crystal and the length of the crystal. The limits of integration would then 
depend on the smaller of these interrelated parameters. If the transmitted 
laser beam is detected after passing through a quarter wave plate and
a crossed 
analyzer, then the  
transmitted intensity I(t) is given by

\be
I(t) = I_0 [\eta + \sin^2(\Gamma_0 + \Gamma_b + \Gamma(t))]
\ee

\ni
where  $\eta $ is the intensity extinction coefficient, $\Gamma_0$  
is the residual retardation by the crystal in the absence of the electric 
field and $\Gamma_b$ is the retardation introduced by the quarter wave plate. 
Typical values for $\Gamma_0$ and $\Gamma(t)$ are in the range of 
tens of milliradians. The value of {\bf $\Gamma $} is chosen to suit the 
detector capabilities and the experimental conditions. 



\section{Measurement of sub picosecond electron bunch length}

This scheme has been successfully used to measure the bunch length of 45 MeV 
electron beam~\cite{semertzidis00,tsang01}. The limit on the
resolution had been the bandwidth of  
the detection system. A number of schemes to measure sub-picosecond electron 
bunch have been proposed so far. These include characterizing the frequency 
modulation on a laser spectrum caused by the electron bunch, performing 
autocorrelation measurements and using the FROG technique to determine
both the  
frequency and time distribution of the laser beam transmitted through a 
birefringent crystal. In the following section another scheme that converts 
temporal information to spatial information to measure sub-picosecond 
electron bunch is described. Bunch length measurement with resolution down 
to the response time of the crystal is possible using this method since 
linear arrays with small pixel dimensions is readily available. Furthermore, 
where pixel dimension proves to be the limitation, effective use of optical 
imaging can be used effectively to overcome this limitation.

A short laser pulse polarized in the YZ plane, 45$^o$ to the z-axis, focused 
using a cylindrical lens to form a line focus, propagates along the y-axis. 
A thin birefringent crystal with  optic axis along z and 
ordinary axis along x is positioned at the waist of the laser beam. The 
electron bunch propagates simultaneously along the x-axis, at a
minimum distance r 
from the laser beam. The transmitted intensity is passed through a crossed 
analyzer and detected by a linear detector array. As shown in the Figure 1, 
only those sections of the laser beam that are below the electron bunch will 
experience a phase retardation linearly proportional to the charge density of 
the electron beam and reach the linear array. The acceptable jitter 
between the electron beam bunch and the laser beam is determined by the x 
dimension of the crystal, length and sensitivity of the detector array, 
length of the line focus of the laser and laser energy available. For a 
typical diode array of 1024 elements, 1 cm crystal length and $\sim $ 100 pJ 
of laser energy in a 1 cm line focus, jitter up to 30 ps can be tolerated as 
well as measured using this arrangement.

\begin{figure}
\includegraphics[width=5.in]{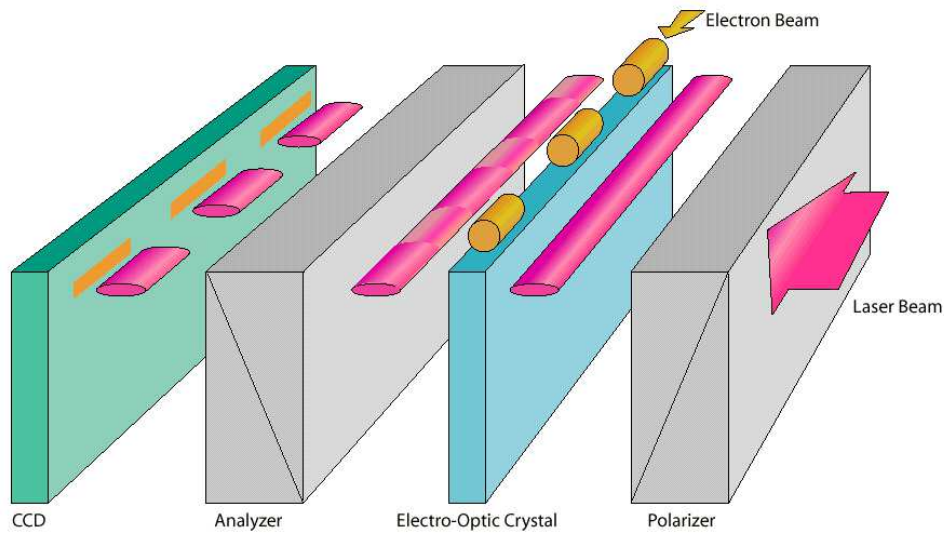}
\caption{Schematic of the experimental arrangement to measure 
subpicosecond electron bunch.}
\label{tr}
\end{figure}

The pulse duration and the thickness (y dimension) of the crystal determine 
the resolution and fidelity of the temporal profile. The distance traveled 
by the electron bunch during the laser pulse constitutes the uncertainty in 
the bunch length measurement. A relativistic electron travels $30 \rm \mu m $ 
during a laser pulse duration of 50~fs (for $n=2$), causing a
corresponding broadening of  
the image on the detector array. The laser pulse duration should then be a 
small fraction of the electron pulse duration to be measured. Short laser 
pulses down to tens of femtoseconds are readily achievable. However, the 
optical beam transport must be designed carefully to reduce pulse 
broadening and minimize high order dispersion.

The phase rotation of a single photon traveling along the crystal is caused 
by the integrated charge density along the diagonal of the sheet of
charge, seen by the photon while in the crystal. Thus, only an  
infinitely thin crystal would preserve the temporal profile of the electron 
bunch. The choice of the thickness of the crystal is, hence, a function of 
the magnitude of the obtainable electric field (determined by charge density 
achievable and distance between the electron beam and the laser), the 
 electro-optic coefficients, the sensitivity of the detection 
system, and the structural integrity of the system.

\bigskip

In conclusion a number of electro-optic detection schemes are
available to measure the  
length of subpicosecond electron bunches. These techniques need to be tested 
for limitations before a judicial choice can be made.

%
%

%
%




\end{document}